\documentclass[a4paper,fleqn,usenatbib]{mnras}

\usepackage[T1]{fontenc}
\usepackage{ae,aecompl}

\usepackage{graphicx}	
\usepackage{amsmath}	
\usepackage{amssymb}	

\def\msun{\hbox{M$_\odot$}}
\title[The outer region of NGC\,6779]{Extra-tidal structures around the {\it Gaia Sausage} candidate globular cluster
NGC\,6779 (M56)}

\author[A.E. Piatti \& J.A. Carballo-Bello]{
Andr\'es E. Piatti$^{1,2}$\thanks{E-mail: andres@oac.unc.edu.ar} and Julio A. Carballo-Bello$^{3}$  \\
$^{1}$Consejo Nacional de Investigaciones Cient\'{\i}ficas y T\'ecnicas, Godoy Cruz 2290, C1425FQB, 
Buenos Aires, Argentina\\
$^{2}$Observatorio Astron\'omico de C\'ordoba, Laprida 854, 5000, 
C\'ordoba, Argentina\\
$^3$Instituto de Astrof\'{\i}sica, Facultad de F\'{\i}sica, Pontificia Universidad Cat\'olica 
de Chile, Av. Vicu\~na Mackenna, 4860, 782-0436, \\ Macul, Santiago, Chile\\
}

\date{Accepted XXX. Received YYY; in original form ZZZ}

\pubyear{2018}

\begin{document}
\label{firstpage}
\pagerange{\pageref{firstpage}--\pageref{lastpage}}
\maketitle

\begin{abstract}
We present results on the stellar density radial profile of the outer regions of NGC\,6779, a
Milky Way globular cluster recently proposed as a candidate member of the {\it Gaia 
Sausage} structure, a merger remnant of a massive dwarf galaxy with the Milky Way.
Taking advantage of the Pan-STARRS PS1 public astrometric and photometric catalogue,
we built the radial profile for the outermost cluster regions using horizontal branch and main 
sequence stars, separately,  in order to probe for different profile trends because of difference 
stellar masses. Owing to its relatively close location to the Galactic plane, we have carefully treated
the chosen colour-magnitude regions properly correcting them by the amount of
interstellar extinction measured along the line-of-side of each star, as well as cleaned them
from the variable field star contamination observed across the cluster field.  In the region spanning 
from the tidal to the Jacobi radii the resulting radial profiles show a diffuse
extended halo, with an average power law slope of -1. While analysing
the relationships between the Galactocentric distance, the half-mass density, the half-light
radius, the slope of the radial profile of the outermost regions, the internal dynamical evolutionary
stage, among others, we found that NGC\,6779 shows structural properties similar to those of 
the remaining  {\it Gaia  Sausage} candidate globular clusters, namely, they are massive clusters  ($>$ 10$^5$$\msun$)
in a moderately early dynamical evolutionary stage, with  observed extra-tidal structures.
\end{abstract} 




\section{Introduction}

Recently, \citet{myeongetal2018} performed a search of Milky Way globular clusters (MW GCs)
that possibly belong to the {\it Gaia Sausage}, an elongated structure in velocity space
created by a massive dwarf galaxy ($\sim$ 5$\times$10$^{10}$ $\msun$) on a strongly radial 
orbit that merged with the MW at a redshift $z \la$ 3 \citep{belokurovetal2018}.
They listed NGC\,1851, 1904, 2298, 2808, 5286, 6779, 6864 and 7089 as probable candidate GCs, 
and NGC\,362 and 1261 as possible ones.

Because of the merger event that gave rise to the {\it Gaia Sausage}, these GCs are expected
to have evidence of strong tidal interactions, such as long tidal tails, azimuthally 
irregular stellar halos, clumpy extended structures, etc. Indeed, 8 out of the 10 candidate GCs 
have previous studies of
their outer regions and all of them show some of the above mentioned signatures. For instance,  
\citet{carballobelloetal2018} found tidal tails around NGC\,1851, 1904, 2298 and 2808;
\citet{carballobelloetal2012} maped the extended envelopes of NGC\,1261 and 6864;
 \citet{vanderbekeetal2015} and \citet{kuzmaetal2016}  detected extra-tidal structures in NGC\,362
and NGC\,7089, respectively.

As far as we are aware, NGC\,6779 (M56) has not been targeted for any analysis of its stellar
structure beyond its tidal radius. However, given that almost all the {\it Gaia Sausage} candidate GCs exhibit extra-tidal features, one would
also expect to find some sort of structures around NGC\,6779. Precisely, the main aim
of this work consists in tracing for the first time the stellar density radial profile of the outermost
cluster regions, and assess on that resulting profile the membership of NGC\,6779 to
{\it Gaia Sausage}. For the sake of the reader, we list in Table~\ref{tab:table1} the adopted values for some
pertinent cluster astrophysical properties.

The paper is organised as follows: In Section 2 we describe the observational data set used
and the reddening corrections performed in order to get intrinsic magnitudes and colours. 
Section 3 deals with the construction of stellar radial profiles for cluster horizontal branch
(HB) and main sequence (MS) stars, respectively, while in Section 4 we analyse and discuss the
resulting radial profiles. Finally, Section 5 summarises the main conclusions of this work.

\begin{table}
\caption{Astrophysical properties of NGC\,6779.}
\label{tab:table1}
\begin{tabular}{@{}llc}\hline
Parameter   &   Value          & Ref. \\\hline
True distance modulus & $m-M_o$ = 15.2$\pm$0.1 mag &  3,4 \\
Heliocentric distance$^a$ & d = 10.96$\pm$0.50 kpc & \\
Core radius           & $r_c$ = 0.44$\arcmin$ (1.40 pc)     &  1,2 \\
Half-light radius     & $r_h$ = 1.12$\arcmin$ (3.57 pc)     &  2 \\
Tidal radius          & $r_t$ = 10.55$\arcmin$ (33.63 pc)    &  1 \\
Jacoby radius         & $r_J$ = (23.40$\pm$1.79)$\arcmin$ (74.60$\pm$5.71 pc) & 6 \\
Ellipticity           &  $\epsilon$ = 0.03 & 1 \\
Mass                  & $M$ = (2.81$\pm$0.52)$\times$10$^5$$\msun$ & 2 \\
Density inside $r_h$  & $\rho_{r_h}$ = 138.0 $\msun$/pc$^3$ &  2 \\
Age                   & $t$ = 12.75$\pm$0.50 Gyr   & 7 \\
Relaxation time       & $t_h$ = 3.1 Gyr & 2\\
Metallicity           & [Fe/H] = -2.0$\pm$0.1 dex &  1,5 \\
\hline
\end{tabular}

\noindent Ref.: (1) \citet{harris1996}; (2) \citet{bh2018};  (3) \citet{sarajedinietal2007}; 
(4) \citet{khamidullinaetal2015}; (5) \citet{carrettaetal2009}; 
(6) \citet{bg2018}; (7) \citet{vanderbergetal2013}.\\

\noindent $^a$ computed from $m-M_o$.
\end{table}

\section{Observational data}

With the aim of looking for extended stellar structures around NGC\,6779,
we  made use of the public astrometric and photometric catalogue produced by
 the Panoramic Survey Telescope and Rapid response System \citep[Pan-STARRS PS1][]{chambersetal2016},
which enabled us to homogeneously cover with deep photometry a large sky area.
We downloaded positions (R.A. and Dec.) and $gr$ PSF photometry for 2365154  stars 
distributed in a box of 3$\degr$$\times$3$\degr$ centred on the cluster. 
We used as quality indicators the errors of the PSF magnitudes gMeanPSFMagErr and
rMeanPSFMagErr to be within the ranges used by \citet{p18a} (see also error bars in 
Fig.~\ref{fig:fig2}).

\begin{figure}
     \includegraphics[width=\columnwidth]{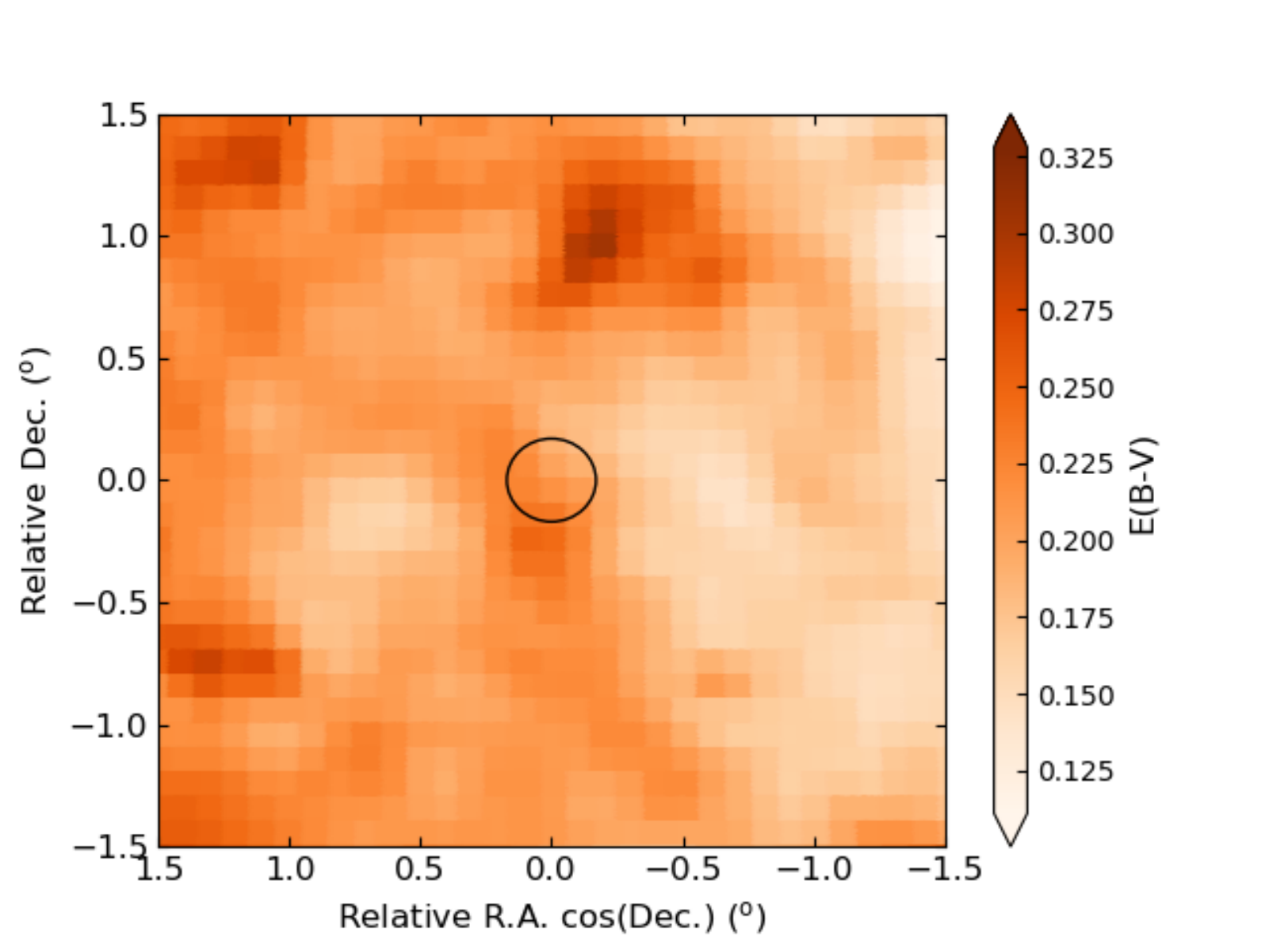}
\caption{Reddening map across the field of NGC\,6779. The circle represents the cluster 
tidal radius (10.55$\arcmin$ (33.63 pc)).}
 \label{fig:fig1}
\end{figure}

\begin{figure}
     \includegraphics[width=\columnwidth]{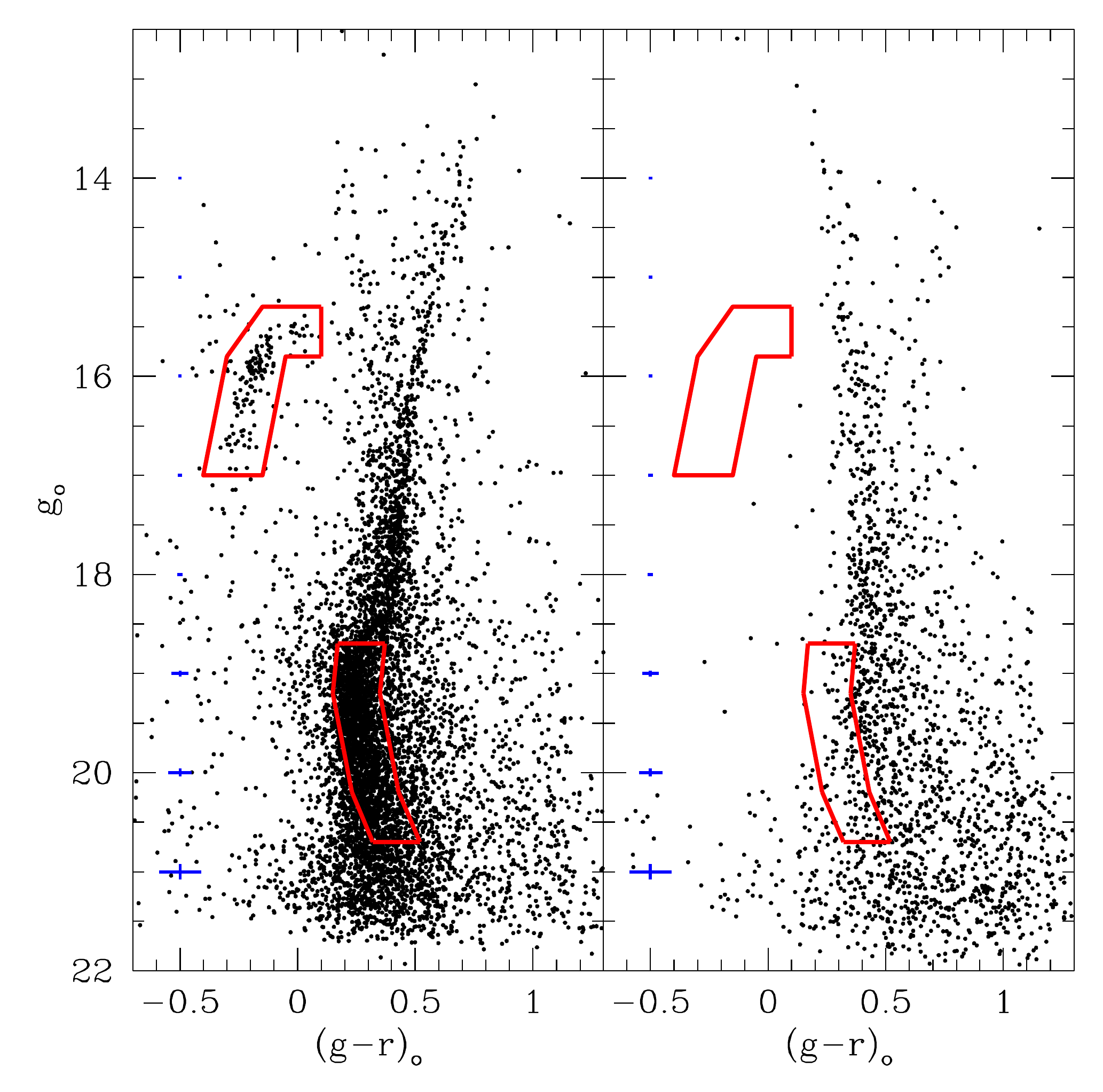}
\caption{CMDs of stars in the field of NGC\,6779 ($r<$ 5; left panel) and in
a MW field with similar area located 1$\degr$ towards the north-west (right panel).
Error bars are included at the left margin of each panel (blue lines). The
two regions used to perform star counts are overplotted with red contour lines.}
 \label{fig:fig2}
\end{figure}

We first examined the spatial variation of the interstellar reddening across
the field containing the cluster. For each entry in our Pan-STARRS PS1 catalogue, we obtained the
$E(B-V)$ value from \citet{sf11} provided by NASA/IPAC Infrared Science 
Archive\footnote{https://irsa.ipac.caltech.edu/}, which is the recalibrated MW extinction 
map of \citet{schlegeletal1998}. We derived a mean colour excess of $<E(B-V)>$ = 0.21$\pm$0.04
mag for the 2365154 stars distributed in the 3$\degr$$\times$3$\degr$ field, with lower 
and upper limits of 0.111 and 0.328 mag, respectively. For a circular region centred on 
NGC\,6779, we obtained $<E(B-V)>$ = 0.20$\pm$0.03 from 31833 stars located within $r_t$, 
with lower and upper values of 
0.139 and 0.249 mag, respectively. This means that this field is affected by
a relative low interstellar absorption, with a slight differential reddening. 
Fig.~\ref{fig:fig1} illustrates the spatial distribution of $E(B-V)$ values.

Then, we derived intrinsic magnitudes $g_o$ and colours $(g-r)_o$ from the Pan-STARRS PS1 
$gr$ magnitudes by correcting them for interstellar absorption, using the individual
$E(B-V)$ values and the $A_\lambda$/$A_V$ coefficients given by \citet{tonryetal2012}.
Aiming at illustrating the wealth of information we gathered,
Fig.~\ref{fig:fig2} depicts the intrinsic colour-magnitude diagram (CMD) of the inner 
cluster region ($r$ $<$ 5$\arcmin$) and that of a sky region with equivalent area located at
1$\degr$  towards the north-west. 

\section{stellar radial profiles}

To trace the stellar density radial profile of the outermost regions of NGC\,6779
we considered cluster HB and MS stars as illustrated
in Fig.~\ref{fig:fig2}, where both groups of stars have been delineated with red
contour lines. HB stars were initially chosen because they are practically not
contaminated by the foreground field, as judged by the absence of them in the CMD of the
comparison star field located at 1$\degr$ to the north-west from the cluster centre. Similar
HB strips result from any selected comparison star field.

\begin{figure}
     \includegraphics[width=\columnwidth]{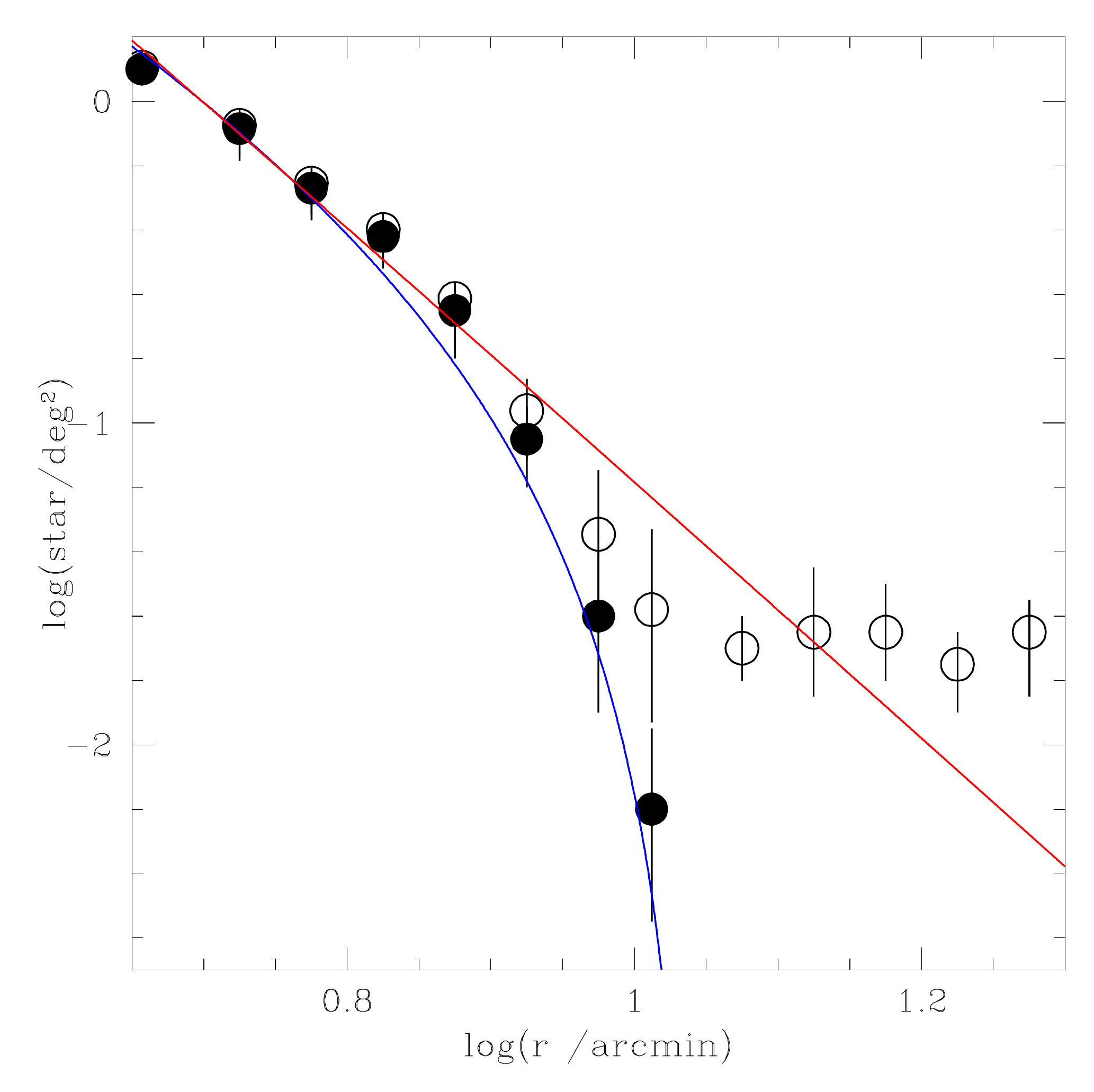}
\caption{Normalised observed and background subtracted stellar density radial profiles with their respective
error bars drawn with open and filled circles, respectively, for HB stars. The curved blue and 
red lines are the models of \citet{king62} and \citet{plummer11}, respectively.}
 \label{fig:fig3}
\end{figure}

We then built the cluster HB density radial profile by applying a kernel density estimator 
(KDE) technique. Particularly, we employed the KDE routine within AstroML \citep{astroml},
which has the advantage of not depending on the bin size and starting point. KDE also 
estimates the optimal FWHM of the Gaussians fitted in a generated grid of 750$\times$750 cells
throughout the 3$\degr$$\times$3$\degr$ field 
to build the stellar density map. The radial profile was then obtained by averaging
the generated stellar density values for annular regions of $\Delta$log(r /arcmin)= 0.05
wide. The resulting stellar density profile is shown in Fig.~\ref{fig:fig3} with
open circles, with the respective error bars. From it, the mean background level was
estimated by averaging those values for log(r /arcmin) $>$ 1.1 
and subtracted from the measured stellar density profile. The background subtracted profile
is depicted with filled circles in Fig.~\ref{fig:fig3}. In this case, the error bars come 
from considering in quadrature the uncertainties of the measured density profile and the 
dispersion of the background level. For comparison purposes, we have overplotted the curves
corresponding to the \citet{king62} and \citet{plummer11} models using the $r_c$, $r_h$
and $r_t$ values of Table~\ref{tab:table1}.

As for the radial profile from MS stars, we defined a strip from the cluster MS turnoff 
down to 2 mags, that expands $g_o$ magnitudes and $(g-r)_o$ colours in the ranges 
(18.7,20.7) and (0.15,0.52), respectively. We decided to go as deep as to be within 100 per 
cent of the photometry completeness; the 50 per cent photometry completeness being at 
$g$=$r$= 23.2 mag, determined with PSF photometry of stellar sources in the stacked images 
(Farrow et al., in preparation). 

\begin{figure}
     \includegraphics[width=\columnwidth]{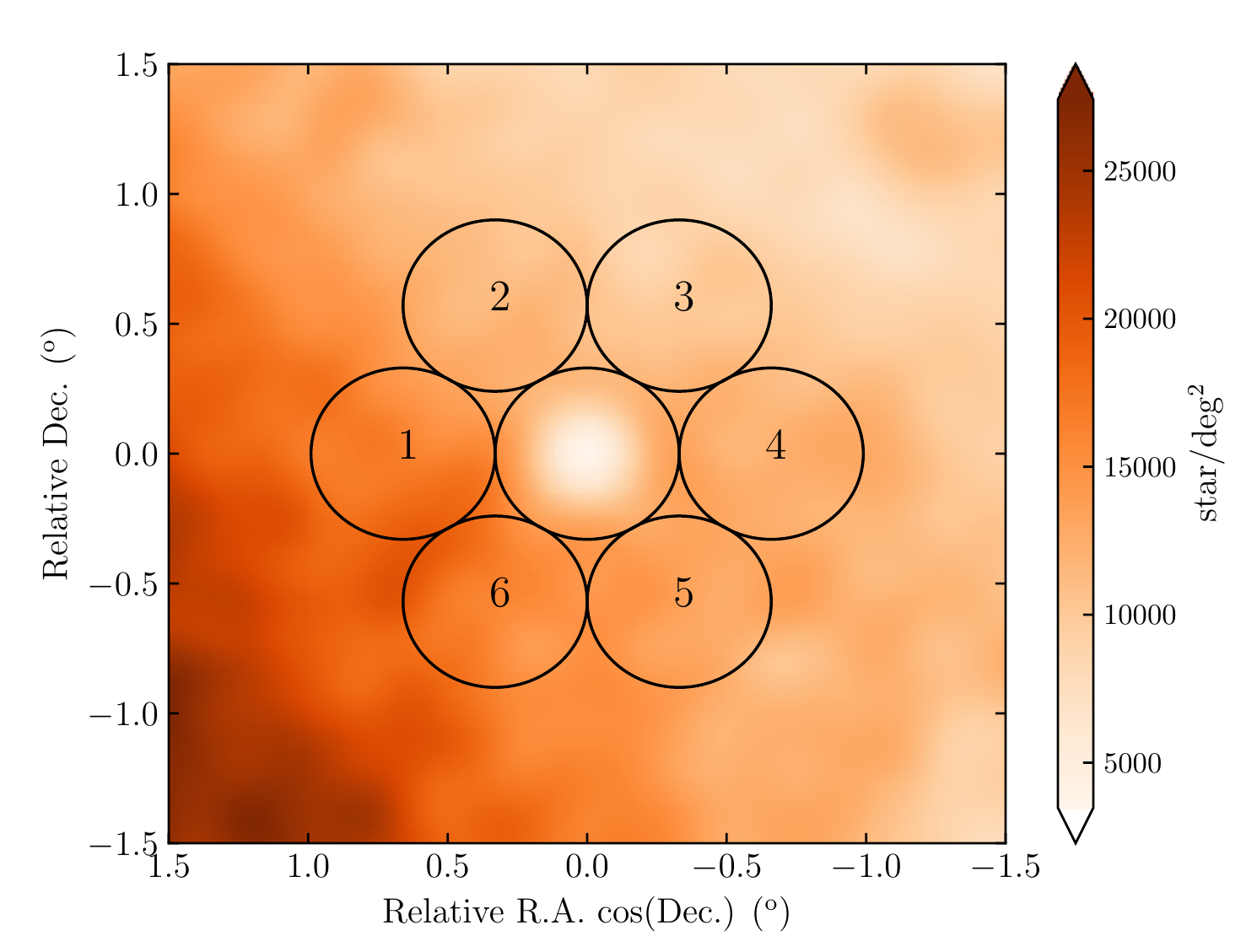}
\caption{MS strip stellar density map of the analysed cluster field with six circular field regions 
overplotted and labelled. We have eliminated the cluster region out to $r_t$ in order to highlight
the variation of the stellar density around it.}
 \label{fig:fig4}
\end{figure}

Unlike the HB strip, the MS one is noticeably
contaminated by field stars, as can be seen in the CMD for the comparison star field
of Fig.~\ref{fig:fig2}. Furthermore, the stellar density and the magnitude and colour distributions 
of stars in that strip vary with the position around NGC\,6779. Indeed,
we built a stellar density map with all these stars using the KDE that visibly reveals 
these variations.  Fig.~\ref{fig:fig4} depicts the resulting density map, where we have 
excluded the inner cluster region ($r<r_t$) in order to highlight the field density inhomogeneities.
The clear density gradient along the south-east north-west direction is due to the position
of NGC\,6779 in the MW ({\it l}= 62$\degr$.66, b=+8$\degr$.34) rather than from reddening
fluctuations. This makes the analysis of the cluster outer regions more challenging, because
we first need to statistically clean the cluster MS strip from the field star contamination 
before building the cluster radial profile. Should we first build the stellar density profile from 
the observed cluster
MS strip stars, would not allow us afterwards to obtain a background subtracted one 
-- as we could satisfactorily do for HB stars --, because the background level around the cluster
is not reasonably uniform for MS strip stars.

In performing the cleaning of the cluster CMD MS strip we considered a circle around the 
cluster centre with radius $r_J$. We also defined 6 different circular star field 
regions distributed around the cluster area of equal size as 
the cluster circle; we have labelled them with numbers 1 to 6 (see Fig.~\ref{fig:fig4}). 
The mean stellar densities of these 6 star field regions turned out to be: 16448$\pm$421, 
11687$\pm$439, 9790$\pm$536,  12085$\pm$349, 13368$\pm$363 and 15762$\pm$261 
star/deg$^2$, 
respectively. Since none of the field areas are representative of that along the line-of-sight 
of the cluster, our strategic approach consisted in decontaminating the cluster MS strip using 
the 6 different reference star fields at a time, and then evaluating the statistical 
significance of any residual structure that might arise from the 6 cleaned CMD MS strips. 
Thus, we compensate cleaning executions using reference star fields less and more dense than 
that along the line-of-sight of the cluster. Notice that we cleaned the cluster area out to its $r_J$. In 
order to extent the cleaning towards farther regions, we would even need to use comparison 
fields located at larger distances from the clustes, which would make in turn the outcomes more
unreliable. Our selected 
reference star fields are located far enough from the  cluster region, but not too far as to 
lose the local distribution in stellar density, magnitude and colour of MW stars.

For each reference star field CMD, we generated a sample of boxes ($g_o$,$(g-r)_o$) 
centred on each star, with sizes ($\Delta$$(g)$,$\Delta$$(g-r)$) defined in such a way that
one of their corners coincides with the closest star in that CMD region.
This procedure of representing the reference star field CMD like an assembly of boxes
was developed by \citet{pb12} and successfully used elsewhere \citep[see, e.g,][]{p17b,p17c,petal2018}. 
It has the advantage of accurately reproducing the reference star field in terms of
stellar density, luminosity function and colour distribution. This is because the number
of assembled boxes is equal to the number of stars in the reference star field CMD, 
as well as the distribution of the magnitudes and colours of the box centres.

The generated box sample of each reference star field CMD was superimposed at a time
to the cluster CMD and subtracted from it one star per box; that
closest to the box centre. 
The resulting cleaned CMD - one per field star CMD used - 
contains mainly 
cluster members, although some negligible amount of interlopers can be expected. 
If the reference star field does not represent that along the line-of-sight of the cluster,
the resulting cleaned cluster CMD is therefore less representative of cluster intrinsic features.
Because of the variation in the stellar density  around of the cluster area, we expect
some difference between the reference star field and that along the line-of-sight of the cluster,
that can blur  the actual cluster features. For this reason, we 
compared the 6 different stellar radial profiles prior to draw any conclusion about the existence
of extra-tidal cluster tracers (see Section 4).

\begin{figure*}
     \includegraphics[width=\columnwidth]{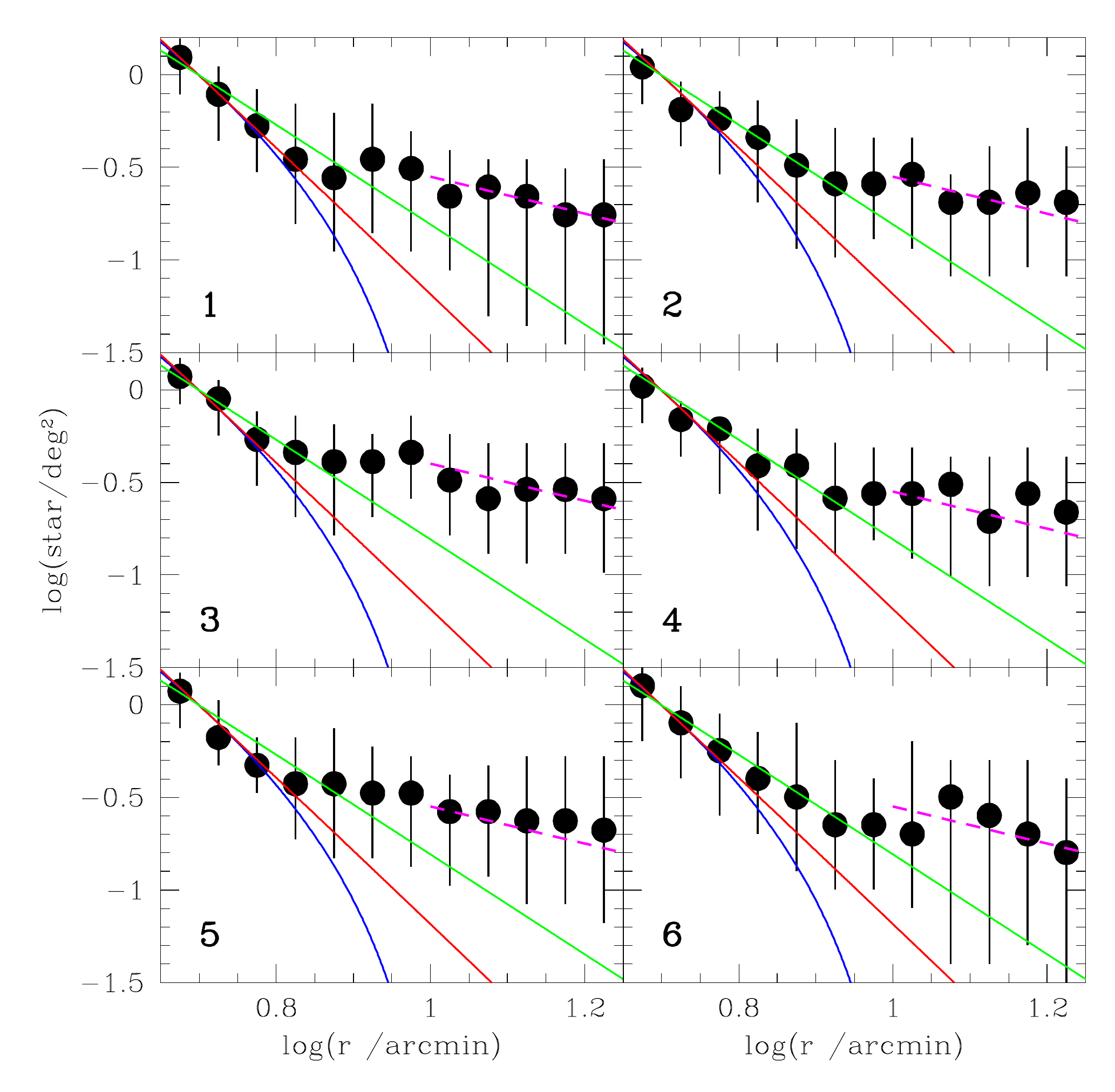}
     \includegraphics[width=\columnwidth]{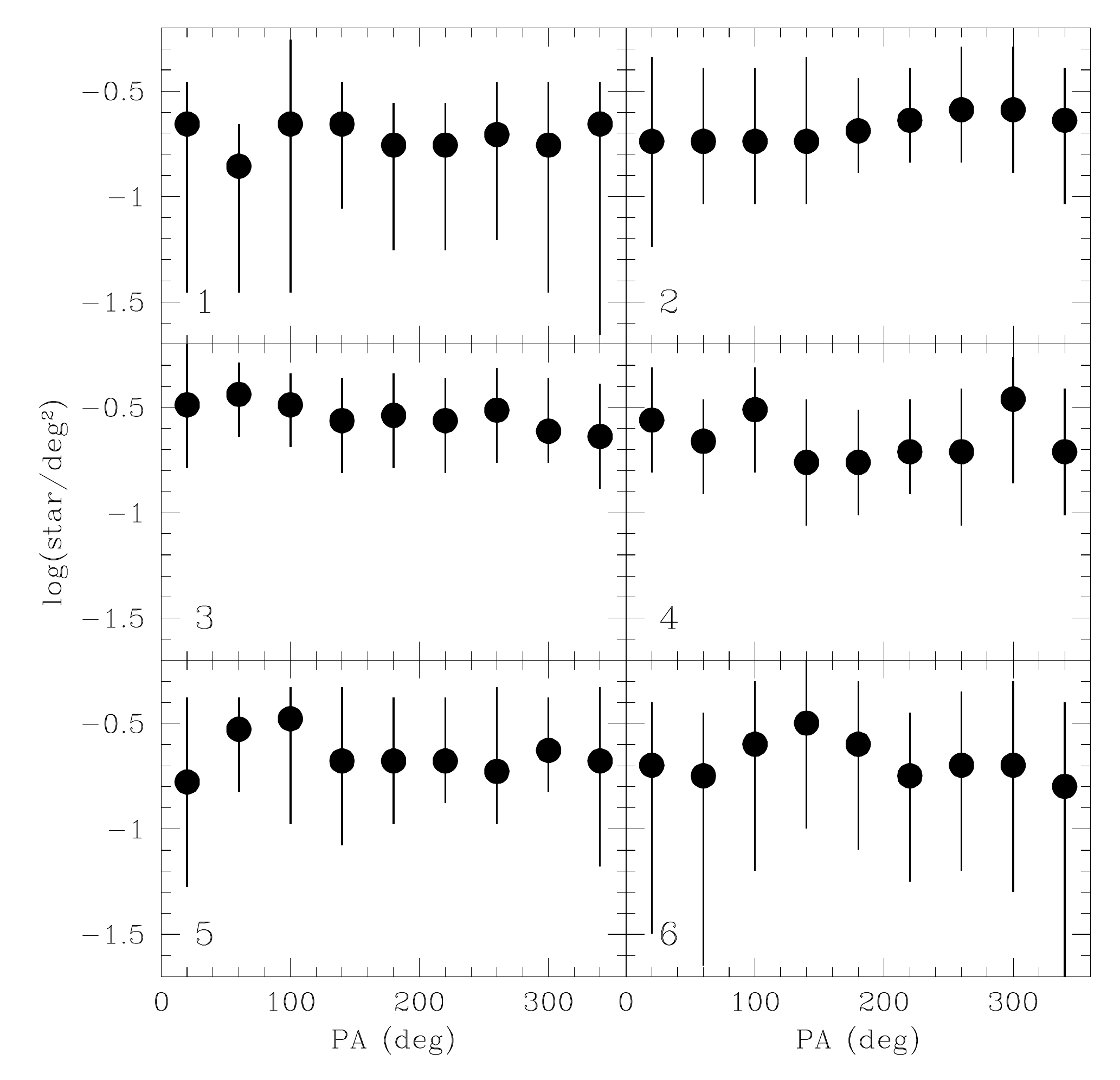}
\caption{{\it Left:} Normalised stellar density radial profiles with their respective error bars 
for MS stars in different decontaminated CMDs, as labelled in each panel
(see also Fig.~\ref{fig:fig4}). The curved blue, red and light-green lines are the models of 
\citet{king62}, \citet{plummer11} and \citet{eff87}, respectively, with $r_c$, $r_h$  and $r_t$ 
taken from \citet{harris1996} and \citet{bh2018}.
The dashed magenta lines represent a power law  $\propto$ $r^{-1}$. {\it Right:} same 
profiles as a function of the P.A. for
stars located farther than $r_t$ from the cluster centre.}
 \label{fig:fig5}
\end{figure*}

The star field cleaned radial profiles were built similarly to that for HB stars, i.e, by 
producing stellar density maps with the KDE and then by averaging all the 
stellar density values inside annuli of $\Delta$log(r /arcmin)= 0.05. For instance,
for the annulus centred at log(r /arcmin)= 0.425, we used 108 values, while at
log(r /arcmin)= 1.225,  we averaged 3540 points. Fig.~\ref{fig:fig5} shows the 6
different resulting radial profiles. Each panel is labelled with the number of the
respective reference star field. We have overplotted \citet{king62}, \citet{plummer11} 
and \citet{eff87} models with blue, red and light-green curved lines, respectively. 
For the \citet{eff87} model, we used the $r_c$ value of Table~\ref{tab:table1} and
our best-fitted value $\gamma$= 2.7$\pm$0.2. Additionally, we overplotted
for the outermost region a power law $\propto$ $r^{-\alpha}$ with a slope $\alpha$ = -1,
using a magenta dashed line. Finally, we repeated the above steps to produce 
azimuthal density profiles for stars located between $r_t$ and $r_J$, also depicted
in Fig.~\ref{fig:fig5}.

\section{Analysis and discussion}

At first glance, Fig.~\ref{fig:fig3} shows that HB stars seem to be tidally filled, with not hint
for extended structures beyond $r_t$. Note that not only the observed radial profile satisfactorily
follows a \citet{king62} model, but also that the adopted $r_c$ and $r_t$ values (see
Table~\ref{tab:table1}) represent very well the cluster HB density radial profile. The lack of evidence
of extra-tidal features from HB stars does not imply that the cluster has not been subjected
to the effects of the MW gravitational potential. As pointed out by \citet{carballobelloetal2012},
stars with smaller masses can be found more easily far away from the cluster main body. For 
this reason, most of the studies devoted to the search for extra-tidal structures have used 
relatively faint MS stars \citep[see, e.g.][]{olszewskietal2009,sahaetal2010,p17c}. Additionally
note that mass segregation also makes more massive stars to be more centrally concentrated
\citep[][and references therein]{khalisietal2007}.

Indeed, the radial profiles constructed in Fig.~\ref{fig:fig5} appear to reveal a different picture of
NGC\,6779. As can be seen, independently of the reference star field adopted to
decontaminate the cluster CMD MS strip, the radial profiles exhibit noticeable excesses 
of stars out to $r_J$. These stellar excesses appear to be important in the context of the
overall cluster stellar density radial profile, as judged by the fact that a relatively  small 
\citet{eff87}'s  $\gamma$ value (2.7) -- very much appropriate to represent extended
halos \citep[see, e.g.][and reference therein]{p18a} -- is not enough to trace the
outermost stellar density profiles. A power law with an average slope equal to -1 is needed
to reproduce the observed trend. At this point, we can conclude 
from the comparison of HB and MS radial profiles, that
there exists a differential mass segregation pattern, being the less massive stars
significanly more segregated.

\citet{gielesetal2011} distinguished from a theoretical point of view tidally affected
(evaporation dominated) from  tidally unaffected (expansion dominated) GCs in the half-mass 
density versus Galactocentric distance plane. Later,  \citet{carballobelloetal2012} reproduced such 
a plot for 114 GCs (see their figure1).  In general, tidally unaffected  GCs are massive objects ($>$ 
10$^5\msun$), among which NGC\,6779 should be included. Indeed, by using a solar
Galactocentric distance ($R_{GC}$) of 8.3 kpc and the cluster heliocentric distance of 
Table~\ref{tab:table1},  we obtained $R_{GC}$ = 10.3$\pm$2.3 kpc, which places the cluster in the 
half-mass density versus $R_{GC}$ diagram into the region of tidally unaffected GCs.
This also happens for all the remaining {\it Gaia Sausage} candidates GCs. Their masses are
in the range (1.2 - 7.4)$\times$10$^5$$\msun$ \citep{bh2018} and their $R_{GC}$ values span from
9.4 up to 18.8 kpc \citet[2010 edition]{harris1996}. However, most of them have been found to
possess tidal tails or diffuse extended structures, thought to be evidence of past tidal
interactions with the MW  (see Sect. 1). If we consider  Pal\,5 and NGC\,5466
(mass $\sim$ (1-4)$\times$10$^4$$\msun$, \citep{bh2018}) -- two low-mass
GCs with evidence of massive tidal tails around them  included in the sample of 
\citet{carballobelloetal2012}, -- we conclude that the distinction between tidally affected
and tidally unaffected GCs is not as straightforward as \citet{gielesetal2011} suggested.
Indeed, recently \citet{webbetal2018} performed controlled $N$-body simulations to systematically
analyse clusters disruption by tidal shocks, and found that the amount of mass lost by the
cluster depends on several factors, among them the strength of the shock, the density of the cluster 
within $r_h$, the number of sub-shocks and the space of time between them.

\citet{carballobelloetal2012} found that the \citet{eff87}'s slope $\gamma$
can be used to discriminate between tidally affected and tidally unaffected GCs.
Those tidally affected GCs not only are low-mass objects, but also have $\gamma$ values
bigger than 4. The typical $\gamma$ value for their tidally unaffected GC sample -- they
show flatter profiles extending to large distances from their compact
cores -- is 3.0$\pm$0.3, in very good agreement with the value derived here for NGC\,6779 
(2.7$\pm$0.2). They also investigated the existence of any relationship of $\gamma$
with the internal structural evolution (dynamical relaxation) and external forces (e.g., 
tidal shocks) using a subsample of 10 GGs with known orbits.
They found that external factors are important in tidally affected GCs, while
internal processes are the main mechanisms of the dynamical evolution of tidally unaffected ones.
From these findings, NGC\,6779 should be currently in an expansion dominated phase.
However, as mentioned above, the \citet{gielesetal2011}'s classification between tidally
affected and tidally unaffected GCs confront with observational evidence. For instance,
NGC\,1851, 1904, 2298 and 2808 are GCs with observed tidal tails and $\gamma$  values
between 2.7 and 3.5 \citep{carballobelloetal2018}.

Stronger tidal fields in the inner parts of the MW can limit the size of GCs. 
\citet{vdberghetal1991} investigated this phenomenon and found a correlation between GC 
$r_h$ values and the respective $R_{GC}$ ones. Recently, \citet{bh2018} confirmed such a 
trend for the half-mass radii of 112 MW GCs. The resulting relationship is far from being tight
(Spearman rank order coefficient of 0.49$\pm$0.07), revealing that tidal effects could
be partially responsible of that correlation. In order to probe this effect, we searched
the \citet[2010 edition]{harris1996} catalogue looking for MW GCs with $R_{GC}$ values similar to that 
of NGC\,6779, within the quoted uncertainty, i.e., 8.0 kpc $<$ $R_{GC}$ $<$ 12.6 kpc. We found 
14 GCs (NGC\,288, 362, 2808, 3201, 4590, 5272, 5286, 6101, 6205, 6341, 7078, 7089, Pal\,11 and 
E\,3) with $r_h$ between 2.1 and 4.8 pc and an average of 4.0$\pm$1.4 pc, in very good 
agreement with the $r_h$ value of NGC\,6779 (3.57 pc). 
All of them are massive objects
(1.2 - 7.4 $\times$10$^5$$\msun$, \citet{bh2018}), while seven have structural analyses
of their outer regions showing a variety of extra-tidal structures, namely:
NGC\,288 presents an extra-tidal clumply structure that extends up to 3.5 times further 
than the cluster $r_t$ \citep{p18a}; NGC\,362 \citep{vanderbekeetal2015}, NGC\,4590, 5272 and 
7078 \citep{carballobelloetal2012} show outer remarkable continuous power-law distributions
extending to large distances from their compact cores; NGC\,7089 exhibits a diffuse
nearly circular-shaped envelope extending to 5 times the nominal cluster $r_t$ value
\citep{kuzmaetal2016} and NGC\,2808 presents tails with different morphologies 
\citep{carballobelloetal2018}. From this result we infer that $R_{GC}$  could mislead our 
interpretation as tidally affected or tidally
unaffected objects. Their own dynamical histories (e.g., the evolution of the eccentricity of their
orbital motions, the number of tidal shocks, the {\it in-situ} or acreeted formation
scenarios) could play a relevant role in shaping their structures.

\begin{figure*}
     \includegraphics[width=\textwidth]{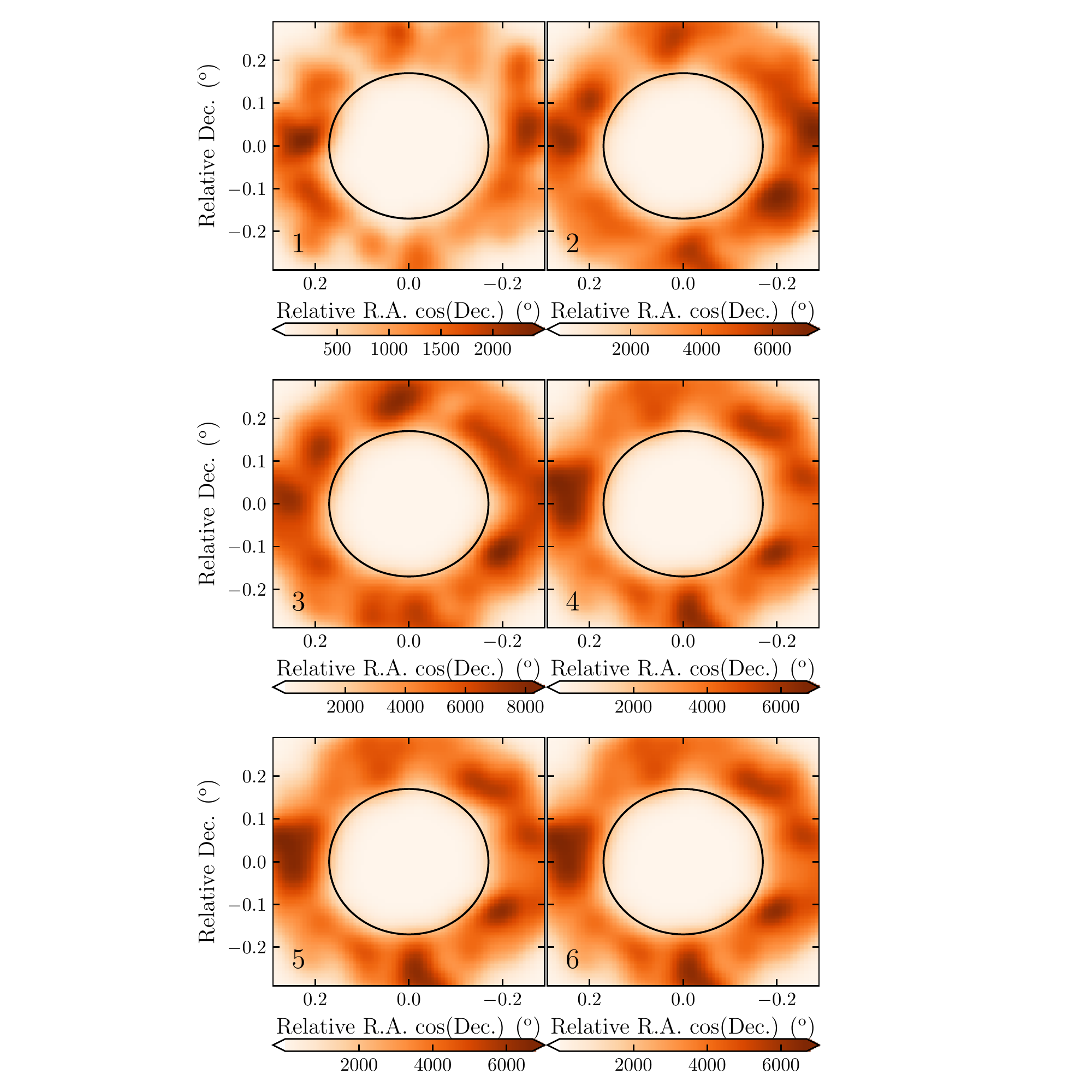}
\caption{MS strip stellar density maps cleaned from field contamination using the
respective six different reference star fields as labelled at the bottom-left corner of each panel
(see Fig.~\ref{fig:fig4}). We have eliminated the cluster region out to $r_t$ in order to highlight
the variation of the stellar density around it. }
 \label{fig:fig6}
\end{figure*}

According to the semi-analytical model proposed by \citet{bg2018} for the evolution of the GC 
mass function in terms of its orbits coupled to a fast stellar stream, the preferencial 
escape of low-mass stars could explain the
absence of tails near massive GCs, despite the fact that massive GCs
lose stars at a higher rate. 
From this model, GCs with
optimal detectability conditions of tidal tails are those with a low remaining
mass fraction $\mu$ -- a measure of its stage of dissolution --
and a high orbital phase $\phi$. The authors highlighted NGC\,6779 as
a candidate to have tidal tails, based on their estimated $\mu$=0.25. However,
this is in tension with the fact that the cluster is a massive GC.
On the other hand, in the case of existing tidal tails,
symmetrically collimated structures should be detected from $r_t$ outwards
\citep[see, e.g.,][]{odenetal2001,belokurovetal2006,noetal2010,sollimaetal2011,balbinotetal2011,erkaletal2017,naverreteetal2017,myeongetal2017,carballobelloetal2018}. 
The right-hand panels of Fig.~\ref{fig:fig5}
do not show a noticeable variation of the radial profiles with the position angle beyond $r_t$, but
evidence of a low-density nearly azimuthally regular envelope.  For the sake of the reader,
Fig~\ref{fig:fig6} illustrates the respective stellar density maps. Note that the appearance of
some clumpy structures depends on the reference star field used.
 
Recently,  \citet{ferraroetal2018} showed that the parameter $A^+$ --  defined as the area 
enclosed between the cumulative radial
distribution of blue straggler stars and that of a reference population --
 is a powerful internal dynamical clock for MW GCs. They obtained $A^+$=0.13$\pm$0.06
 for NGC\,6779,
which implies that the cluster has a modest level of internal evolution. Taking this result
into account we infer that the extended diffuse, nearly azimuthally regular structure found in 
NGC\,6779  could have been caused by the effects of the MW potential, rather than dominated
by internal relaxation, although evidence for
tidal shocks has not been detected. Furthermore, almost all {\it Gaia Sausage} candidates GCs 
  would not seem to be experiencing  an advanced internal dynamical stage either, as judged by their
  estimated $A^+$ values, which span from 0.10 up to 0.25, with an average of 0.18 (6
  GCs); NGC\,1851 being the sole exception ($A^+$=0.48).

\section{Conclusions}

NGC\,6779 is among the 10 MW GCs suggested to belong to the {\it Gaia Sausage}, a 
structure that emerged from the accretion of a massive dwarf galaxy to the MW. Almost all
candidate GC members exhibit some kind of extra-tidal feature (tails, extended halos, etc),
witnesses of the strong tidal interactions that could have taken place during the merger event.
The outermost regions of NGC\,6779 have not been studied so far, so that it is still worth to
analyse them in order to investigate whether they differ from the remaining  {\it Gaia Sausage}
candidate GCs. 

We made use of  the Pan-STARRS PS1 public astrometric and photometric catalogue
to trace for the first time the outer stellar density radial profile of NGC\,6779, reaching
its $r_J$ radius.  In doing that we first corrected the $gr$ magnitudes of each observed star 
located in a field of 3$\degr$$\times$3$\degr$ centred on the cluster according to the
amount of interstellar extinction along the line-of-sight of that star. Fortunately, 
although the cluster is placed at a relatively low Galactic latitude, the reddening map
revealed relatively small $E(B-V)$ colour excesses with a relative low signature
of differential reddening as well. Then, we delineated two
CMD regions, one embracing the cluster HB and another along the MS, from the cluster
MS turnoff down to 2 mag. 

Because of the visible variation of the field star density across the cluster field, we
cleaned both devised regions from field contamination by using a  method that 
gets rid of the actual luminosity function and colour distribution of field stars from
the cluster CMD. We then used the unsubtracted to build stellar density maps from an 
optimised KDE technique, which in turn, were used to construct the cluster radial profiles. 
To clean the cluster CMD we used six different reference star fields uniformly distributed
around the cluster, that span all the star field scenarios, from those less dense than that
along the cluster line-of-sight up to those more crowded. All the resulting radial
profiles show a diffuse nearly continuos structure that extends from the cluster $r_t$
until its $r_J$. The general trend  of this extra-tidal feature can be represented by a
power law with an average slope of -1. This means that the diffuse extended halo is
less steep than our best-fitted \citet{eff87} model ($\gamma$= 2.7), which 
suggest that it could extend even further.

We made use of diagnostic diagrams proposed as good discriminators between
tidally affected and tidally unaffected GCs, such as the Galactocentric distance versus
half-light radius plane and the half-mass density versus Galactocentric distance diagram.
We also investigated the possible origin of the extended halo from the internal
dynamical clock $A^+$ index and the dependance of $\gamma$ with the GC
orbital parameters. 

In general, we found that the link between Galactocentric distances, half-light radius,
half-mass density, present-day cluster mass, orbital parameters, among others, and the origin
of extra-tidal structures is not straightforward as previously thought. There are
a number of issues to be considered along the lifetime of the GCs, such as the occurrence 
of tidal-shocks, the number and the time spacing of them, that can result in different
outcomes (more or less mass lose, spatial pattern of the escaping stars, etc). In the
case of NGC\,6779, we found structural properties that are within the values spanned among the
remaining {\it Gaia Sausage} candidates GCs. They are massive clusters ($>$ 10$^5$$\msun$),
with flatter profiles extending to large distances from their compact cores ($\gamma$ $\sim$ 3.0),
and internal dynamical clock $A^+$ index revealing a moderate evolutionary stage. Precisely,
based on this latter parameter we speculate with the possibility that all the variety of
extended features seen in the {\it Gaia Sausage} candidates GCs -- from now NGC\,6779 also 
included -- represent the footprints of the tidal interaction with the MW.

\section*{Acknowledgements}
We thank the referee for the thorough reading of the manuscript and
timely suggestions to improve it. 
JAC-B acknowledges financial support to CAS-CONICYT 17003.
The Pan-STARRS1 Surveys (PS1) and the PS1 public science archive have been made possible through contributions by the Institute for Astronomy, the University of Hawaii, the Pan-STARRS Project Office, the Max-Planck Society and its participating institutes, the Max Planck Institute for Astronomy, Heidelberg and the Max Planck Institute for Extraterrestrial Physics, Garching, The Johns Hopkins University, Durham University, the University of Edinburgh, the Queen's University Belfast, the Harvard-Smithsonian Center for Astrophysics, the Las Cumbres Observatory Global Telescope Network Incorporated, the National Central University of Taiwan, the Space Telescope Science Institute, the National Aeronautics and Space Administration under Grant No. NNX08AR22G issued through the Planetary Science Division of the NASA Science Mission Directorate, the National Science Foundation Grant No. AST-1238877, the University of Maryland, Eotvos Lorand University (ELTE), the Los Alamos National Laboratory, and the Gordon and Betty Moore Foundation.

\bibliographystyle{mnras}

\input{paper.bbl}







\bsp	
\label{lastpage}
\end{document}